\documentclass[sigplan,10pt]{acmart}

\acmConference[]{}{}{}
\acmYear{2018}
\acmISBN{} 
\acmDOI{} 
\startPage{1}

\setcopyright{none}
\renewcommand\footnotetextcopyrightpermission[1]{} 

\usepackage{booktabs}   
\usepackage{caption}
\usepackage[labelformat=parens]{subcaption} 
                        
\usepackage{algorithm}
\usepackage{algpseudocode}
\algrenewcommand\algorithmicindent{0.9em}

\definecolor{kscolor}{rgb}{0.9,0.1,0.1}

\definecolor{mpcolor}{rgb}{0.1,0.6,0.1}

\definecolor{rbcolor}{rgb}{0.1,0.1,0.9}

\definecolor{todocolor}{rgb}{1,0.6,0.1}

\newcommand\code[1]{\texttt{#1}}

\newcommand{\jsnice}{JSNice}
\newcommand{\jsnaughty}{JSNaughty}
\newcommand{\uglifyjs}{UglifyJS}
\newcommand{\R}{\mathbb{R}}
\newcommand{\name}{\textsc{Context2Name}}

\begin{document}

\sloppy

\title[\name{}]{\name{}: A Deep Learning-Based Approach to Infer Natural Variable Names from Usage Contexts}         

\author{Rohan Bavishi}
\affiliation{
  \department{EECS Department}
  \institution{University of California, Berkeley, USA}
}

\author{Michael Pradel}
\affiliation{
  \department{Department of Computer Science}
  \institution{TU Darmstadt, Germany}
}

\author{Koushik Sen}
\affiliation{
  \department{EECS Department}
  \institution{University of California, Berkeley, USA}
}

\begin{abstract}
Most of the JavaScript code deployed in the wild has been minified, a process in which identifier names are replaced with short, arbitrary and meaningless names.
Minified code occupies less space, but also makes the code extremely difficult to manually inspect and understand.
This paper presents \name{}, a deep learning-based technique that partially reverses the effect of minification by predicting natural identifier names for minified names.
The core idea is to predict from the usage context of a variable a name that captures the meaning of the variable.
The approach combines a lightweight, token-based static analysis with an auto-encoder neural network that summarizes usage contexts and a recurrent neural network that predict natural names for a given usage context.
We evaluate \name{} with a large corpus of real-world JavaScript code and show that it successfully predicts 47.5\% of all minified identifiers while taking only 2.9 milliseconds on average to predict a name.
A comparison with the state-of-the-art tools \jsnice{} and \jsnaughty{} shows that our approach performs comparably in terms of accuracy while improving in terms of efficiency.
Moreover, \name{} complements the state-of-the-art by predicting 5.3\% additional identifiers that are missed by both existing tools.
\end{abstract}

\ccsdesc[500]{Software and its engineering~General programming languages}
\ccsdesc[300]{Social and professional topics~History of programming languages}


\maketitle


\section{Introduction}

Developers invest a significant portion of their time in reading and understanding code~\cite{Singer1997}. This is because programmers need to periodically review their code for defects, to look for regions of code to optimize, extend existing functionality or simply increase their knowledge base~\cite{Bacchelli2013}. The developer community has come up with various guidelines and styles to be followed while writing programs that can potentially reduce the comprehension overhead. One widely accepted guideline is to use meaningful variable and function names. Ideally, a name should capture its semantic function in a program, effectively acting as an abstraction that developers can use to aid their understanding~\cite{Gellenbeck1991}.

In the same vein, however, variable and function names when deliberately designed poorly, can provide a layer of obfuscation that discourages review and inspection. Combined with the removal of formatting, such as indentation and white-spaces, while retaining functionality, can make a program extremely difficult to read for a developer.
Code with non-meaningful identifier names is particular common for real-world JavaScript, where most developers apply \emph{minification} before shipping their code.
This process replaces all local identifier names with short, arbitrary, and meaningless names.
A number of publicly available tools automate minification by \emph{mangling} local names into short, cryptic ones, and by aggressively reusing names in different scopes.
The resulting JavaScript files are smaller and thus reduce the download time of, e.g., client-side web application code.
In addition, some website might wish to conceal the meaning of the code to protect their intellectual property or to hide their malicious intent.

Minification tools usually produce \emph{source-maps} that map various elements in the minified code back to their original counterparts. Unfortunately, in most cases source-maps are available only to the authors of the JavaScript code as they are not shipped along with the source. To enable external reviewing and security analysis, it is important to develop techniques that attempt to recover the original source code from its minified version, primarily by renaming identifiers to more meaningful names.

This paper addresses the challenge of inferring natural variable names in minified code through deep learning.
The key idea in our approach is to capture the syntactic usage context of a variable or function across a JavaScript program and to predict a natural name from this context.
To gather usage contexts, we use a lightweight technique that views code as a sequence of lexical tokens and extracts sequences of tokens surrounding each use of a variable.
We then use these sequences to train a \emph{recurrent neural network} (RNN)~\cite{LSTM1997} that predicts a natural identifier for the given usage context.
Since these sequences can be arbitrarily long, we use a separate \emph{auto-encoder} neural network to generate \emph{embeddings}, which are much smaller in dimensionality, and retain the key features that are sufficient to categorize usage contexts.
The RNN used for prediction can be trained much more efficiently with these embeddings.

To train our \name{} approach, we leverage the huge amount of JavaScript code that is available online. Machine learning techniques exploiting this availability of source code have been used to solve a variety of development tasks, such as code completion~\cite{Raychev2014}, fixing syntactic errors~\cite{Gupta2017}, code clone detection~\cite{White2016}, malware analysis~\cite{David2015}, and even generating programs with specific constraints~\cite{Shu2017}. Deep Learning~\cite{LeCunn2015DPL} is a fast-growing field in machine learning that has been very successful in natural language tasks, such as completion, translation and summarization.
We show that this effectiveness propagates to the problem of inferring meaningful variable names as well.

We evaluate our technique on a large, publicly available corpus of JavaScript source code, wherein \name{} is able to recover 47.5\% of all minified identifiers \emph{exactly}, i.e., our tool predicts the \emph{same} name that the authors of the original programs had in mind.
We also show that our approach is practical. It takes an average of 2.9 milliseconds to predict a name, or 110.7 milliseconds to process a file, on average.

A comparison with the state-of-the-art tools \jsnice{}~\cite{Raychev2015} and \jsnaughty{}~\cite{Vasilescu2017} shows that our work achieves comparable accuracy while improving in efficiency.
Evaluating all tools on the same data set, along with a reasonable time limit for processing, \name{} is able to recover about as many identifiers as \jsnice{} and 8.1\% more identifiers than \jsnaughty{}.
The main contribution of our work is the conceptually much simpler approach that will be easier to adapt to another programming language than \jsnice{}.
The reason is that our work does not perform any program analysis for feature extraction, but instead relies in neural networks to learn which parts of a token stream are relevant for predicting natural names.

In summary, this paper contributes the following:
\begin{itemize}
    \item A deep learning-based framework to recover natural identifier names from minified JavaScript code.
    
    \item A technique for computing vector embeddings of the usage contexts of a variable. The technique makes minimal assumptions about the underlying programming language, and can be adapted to other languages and usage scenarios.
    
    \item Empirical evidence that the approach exactly recovers 47.5\% of all minified names in a large, publicly available corpus of JavaScript code, which is comparable to existing approaches. The average time for processing each file is also well under a second.
    
\end{itemize}

The implementation of our approach, as well as all data to reproduce our results, will be made available as soon as the paper gets accepted.



\begin{figure*}[tb]
  \centering
  \includegraphics[width=\textwidth]{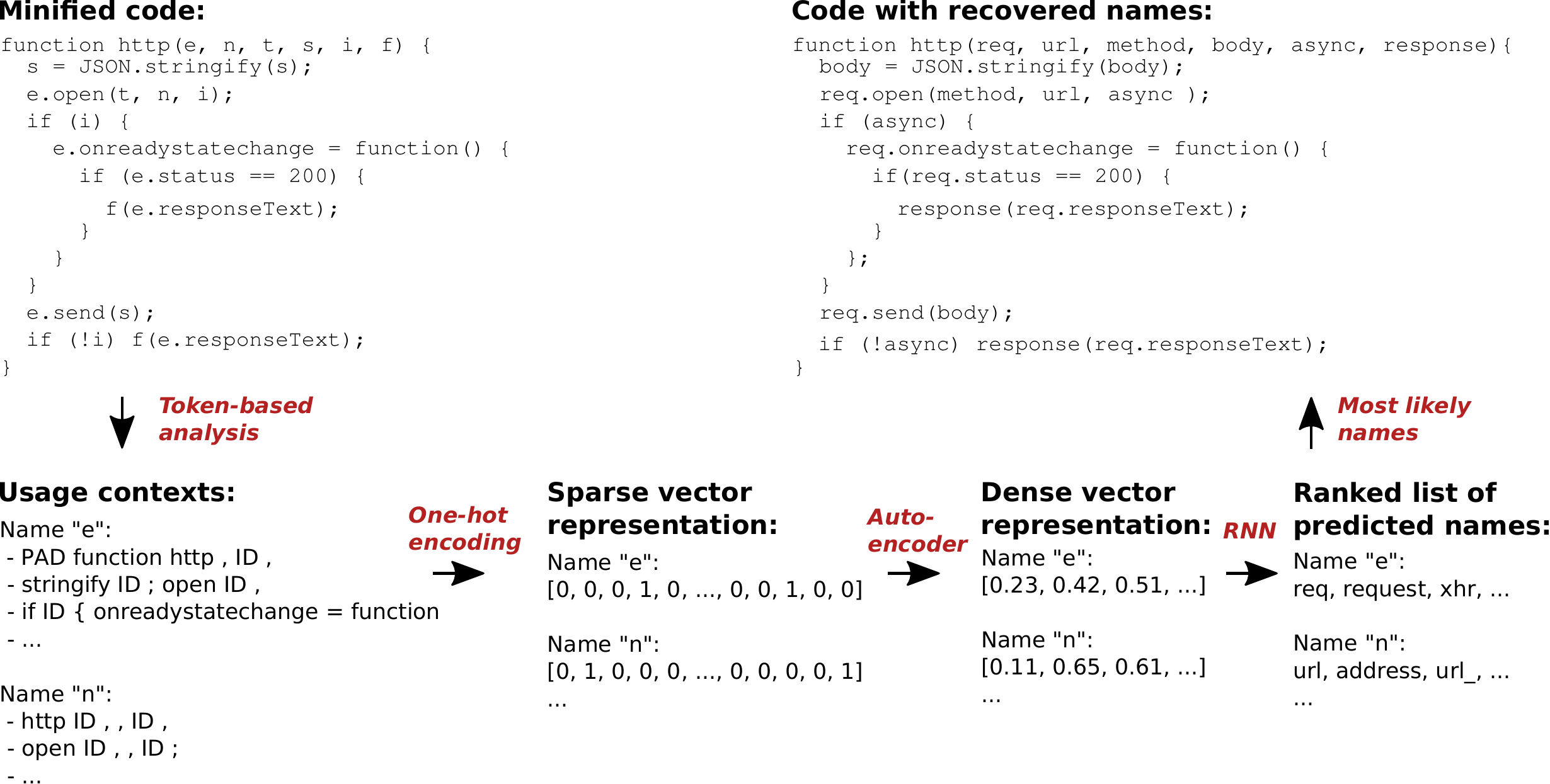}
  \caption{Overview of the \name{} approach.}
  \label{fig:overview}
\end{figure*}

\section{Overview and Example}
\label{sec:overview}

This section presents an informal outline of our approach and illustrates the main ideas with an example.
The example is a piece of JavaScript code, shown in the upper-left corner of Figure~\ref{fig:overview}.
The code has been minified with \uglifyjs{}, a popular minification tool that replaces all local variable names with short, meaningless, and arbitrarily chosen names.
All globally visible names, such as \texttt{http}, and property names, such as \texttt{responseText}, are not minified to ensure that the code preserves the semantics of the original code.
In practice, \uglifyjs{} also removes unnecessary white-space and indentation, which we preserve here for clarity.

The minification makes it difficult to understand the meaning of individual variables and the overall purpose of the code.
The goal of \name{} is to recover meaningful and natural names that help developers not familiar with the code to reason about it.
Ideally, the recovered names are those chosen by the developers in the original code.
In general, predicting the exact same names as in the original code is not always possible.
The reason is that, even though developers generally tend to write regular and ``unsurprising'' code~\cite{Hindle2012}, developers do not always choose the most natural identifier name, but sometimes choose a semantically equivalent name or a slightly misleading name.
To achieve the goal of helping developers understand the code despite minification, it is therefore sufficient to recover some meaningful name, not necessarily the original name, for each local variable.

For the example, the upper-right corner of Figure~\ref{fig:overview} shows the code with names recovered by \name{}.
Five out of the six local names have been recovered exactly, i.e., as in the original code: \texttt{req}, \texttt{url}, \texttt{method}, \texttt{async}, and \texttt{body}.
The remaining name, \texttt{response}, is also very similar in spirit to the name \texttt{callback} that was used by the author of the original code.
Overall, the recovered code very much reflects the developer's intention, making it easy for another developer to reason about the code.

The key insight used by our approach to recover names is that the code surrounding different occurrences of a local variable offers a lot of information about the meaning of the variable.
For example, the tokens surrounding \texttt{e} in Figure~\ref{fig:overview} include the keyword \texttt{function}, the global function name \texttt{http}, and the property names \texttt{open}, \texttt{oneadystatechange}, \texttt{send}, and \texttt{responseText}.
An average JavaScript developer can infer from this information that \texttt{e} is likely to be an \texttt{XMLHttpRequest} object.
A more difficult case is the minified name \texttt{i}.
Its surrounding tokens include \texttt{if}, \texttt{onreadystatechange}, and \texttt{responseText}, which at first glance do not reveal much about the purpose of \texttt{i}.
A skilled developer who has used \texttt{XMLHttpRequest} may learn from the fact that \texttt{i} is the third argument of \texttt{open} and that \texttt{onreadystatechange} is close to the \texttt{if} condition where \texttt{i} evaluates to true, while \texttt{responseText} is close to another \texttt{if} condition that involves \texttt{i} and a negation.
All these features suggest that \texttt{i} toggles the asynchronous behavior of an HTTP requests.

Our technique essentially simulates this semantic reasoning based on prior knowledge of JavaScript code.
However, instead of relying on an expert JavaScript developer, we exploit the power of advanced machine learning techniques and the availability of large corpora of code.
The approach identifies regularities in the way identifier names occur in real code and uses these regularities to predict meaningful names.

Figure \ref{fig:overview} illustrates the different steps taken by \name{} during this process:
\begin{enumerate}
    \item At first, a simple token-based analysis of the code extracts usage contexts for every minified name.
    Specifically, the analysis extracts one usage context for each occurrence of the name.
    The usage context consists the sequence (of length 3 in this case) of lexical tokens before and after the minified name.
    \item The next step converts the set of usage contexts of a name into a vector representation.
    This representation is based on a standard one-hot encoding and yields a sparse, binary vector, where most elements are zero and only few elements are one.
    \item To ensure the scalability and efficiency of the overall approach, the next step converts the sparse vectors into a dense vector representation.
    This step is fully automated by using a sequence auto-encoder~\cite{Dai2015}, i.e., an semi-supervised neural network that searches an efficient encoding.
    \item The dense vector representations are fed into a supervised machine learning model, a recurrent neural network (RNN).
    We train this model to predict a ranked list of meaningful names for a given vector representation of the usage contexts of a variable.
    A key insight is that the prediction depends only on the usage contexts of the variable, and not on the minified name.
    \item Finally, \name{} greedily selects for each minified variable the name predicted with maximum probability and outputs the code with recovered names.
\end{enumerate}

An important conceptual benefit of \name{} over the state of the art technique \jsnice{} is that our approach makes very little assumptions about the language of the analyzed programs.
Concretely, our approach abstracts program code into a sequence of tokens and assumes to have a way to identify occurrences of the same local variable.
In addition to these assumptions, JSNice extracts various hard-coded kinds of relations between program elements from the code, such as between different identifiers involved in the left-hand side and right-hand side of assignments.
These relations are specific to the analyzed language, making it non-trivial to adapt JSNice to a different programming language.
Instead, \name{} leaves the task of identifying relevant structural relations between program elements to a machine learning model, making it easier to adapt the approach to other languages.

\section{Approach}

In this section, we provide the technical details of our approach.
The key idea is to approximate the semantic meaning of a variable or function using a sequence of lexical tokens surrounding its different points of usage in the code.
We formally describe this notion of a usage summary in Section~\ref{sec-usage-sequence}.
Sections~\ref{sec:autoEncoder} and~\ref{sec-predict} then describe two neural networks: 
One network to reduce the usage summary into an efficient vector representation, and another network to predict a suitable name for a given usage summary.
Finally, Section~\ref{sec:recovery} presents how to recover all minified names of a program while preserving the semantics of the code.

\subsection{Extracting Usage Summaries}
\label{sec-usage-sequence}

The input to \name{} is the source code of a minified program.
As a first step, the approach is to extract a usage summary for a each local variable or locally defined function.
The usage summary will be used by later steps of the approach to predict a suitable name for the variable.
A usage summary is composed of the different contexts in which an identifier is used.
To extract the usage summary, we view the source code as a sequence of lexical tokens $\mathcal{T} = <t_0, t_1, \cdots , t_{len}>$.
We drop some tokens, specifically dot punctuators and round parentheses, from $\mathcal{T}$ as we did not find them to be particularly useful.
Let $\mathcal{N} = \{n_1, n_2, \cdots , n_k\} \subseteq \mathcal{T}$ be the set of all local names in the code.
Because a single name may independently occur in multiple scopes, variables or functions in different scopes having the same name have separate entries in $\mathcal{N}$.

For constructing the context of each occurrence of an identifier, we define a helper function.
Let $\mathcal{T}[k]$ denote the $k^{th}$ token in the sequence $\mathcal{T}$. We define a token projection function $g_\mathcal{T}$ for $\mathcal{T}$ as follows:
\begin{equation}
    g_\mathcal{T}(k) = \begin{cases}
               \ \ \mathcal{T}[k] \text{ if } 0 \leq k \leq len \land \mathcal{T}[k] \notin \mathcal{N}\\
               \ \ \text{\texttt{ID}} \text{  if } 0 \leq k \leq len \land  \mathcal{T}[k] \in \mathcal{N}\\
               \ \ \text{\texttt{PAD}} \text{  if } k < 0\\
               \ \ \text{\texttt{PAD}} \text{  if } k > len
             \end{cases}
\end{equation}
The function replaces all local names in the code with a special \texttt{ID} token.
The reason is that these names are minified in the given code and therefore do not contribute any semantic information.
Also note that $g_\mathcal{T}$ returns the special padding token \texttt{PAD} when its argument is out of range.
This case is useful for extracting the context of tokens that appear close to the beginning and end of the analyzed token sequence $\mathcal{T}$.

Based on the helper function $g_\mathcal{T}$, we now define the context of an occurrence of a local name.
For each occurrence of a local name in $\mathcal{T}$, we extract the $q$ preceding tokens and the $q$ following tokens into a the \emph{\textbf{context}} $c(t)$ of a token $t$ in~$\mathcal{T}$:
\begin{align}
\begin{split}
    c(t) = < & g_\mathcal{T}(k-q), \cdots, g_\mathcal{T}(k-1), \\
           & g_\mathcal{T}(k+1), \cdots, g_\mathcal{T}(k+q)> 
            \text{if } t = \mathcal{T}[k]
\end{split}
\end{align}
This local context captures the usage of a name at a particular code location.
The hyper-parameter $q$  can be configured to adjust the number of tokens extracted as context. We use $q=5$ as a default in our evaluation.

Finally, for each local name $n$, we concatenate the contexts for different usages of $n$ in the code into a single sequence of tokens, which we call the \emph{\textbf{usage summary}} of the name.
We use up to $l$ contexts per name to construct the usage summary, where $l$ is another hyper-parameter that can be configured to adjust the size of summaries.
If less than $l$ contexts are available for a particular name, we pad the sequence with the special \texttt{PAD} tokens. If more than $l$ contexts are available, we use the first $l$ contexts. We use $l=5$ as a default in our evaluation.
We formally define the usage summary as a function $\mathcal{U}(n)$ where $n \in \mathcal{N}$ as follows ($\circ$ is the sequence concatenation operator):

\begin{align}
\begin{split}
     \mathcal{U}(n) = &\  c(t_1) \circ c(t_2) \circ \cdots \circ c(t_{l}) \\
                  & \text{where } t_i \ \ \forall 1 \leq i \leq l \ \ \text{are occurrences of} \ n
\end{split}
\end{align}


For our running example from Section~\ref{sec:overview}, the lower left corner of Figure~\ref{fig:overview} show the usage contexts of variables \code{e} and \code{n} (with parameter $q=3$ for space reasons).

The set of tokens in the usage contexts, along with their position relative to the variable occurrences, captures the syntactic context of the variable usages.
The intuition behind \name{} is that this context is often sufficient to infer the meaning of the variable.

\subsection{Learning Embeddings for Usage Summaries}
\label{sec:autoEncoder}

After extracting usage summaries for each local name in the code, the next step is to summarize them into an efficient vector representation called \emph{embedding}.
The motivation for this step is twofold.
First, to benefit from a machine model that predicts likely variable names, we need to convert the information extracted from source into a format suitable for machine learning.
The neural network model we use here, as many other machine learning models, expects vectors of real numbers as its input.
Second, the usage summaries are highly redundant, e.g., because the same kind of token occurs many times and because subsequences of tokens occur repeatedly.
To ensure the scalability and efficiency of the overall approach, we compress usage summaries into an efficient vector representation.

One option to convert usage summaries into a compact vector representation would be to manually define a set of features and to create vectors that describe the presence or absence of these features.
However, coming up with a suitable set of features that capture the semantics of identifier usages in JavaScript would require significant manual effort.
Moreover, manually designed features would tie our approach to a specific programming language, and require additional manual effort to adapt it to another language.

Instead of manually defining how to compress the usage summaries, we use an auto-encoder.
An auto-encoder is a supervised neural network model that learns to compress a given vector while preserving as much of the original information as possible.
We train a sequence auto-encoder~\cite{Dai2015} that compresses each context in a usage summary into a compact vector, allowing us to represent the usage summary as the concatenation of these compact vectors.

The first step is to define an input vocabulary $\mathcal{V}_{inp}$ of tokens that the network recognizes. We construct the input vocabulary by picking the most frequent $|\mathcal{V}_{inp}|$ tokens (plus the special \texttt{PAD} token) across our training set of code.
In our experiments, $|\mathcal{V}_{inp}| = 4,096$.
All tokens that are not frequent enough to occur in $\mathcal{V}_{inp}$ are represented by a special \texttt{UNK} token.
The input vocabulary is then used to convert the tokens in the usage summary to their one-hot representations, i.e., binary vectors of size $|\mathcal{V}_{inp}|$ with the $i^{th}$ bit set for the $i^{th}$ token in $\mathcal{V}_{inp}$.
The size $s_c$ of the one-hot representation of the context $c(t)$ of a token $t$ is $s_c = |\mathcal{V}_{inp}| \cdot q \cdot 2$ because the context contains $q$ tokens before and $q$ tokens after $t$, each represented by $|\mathcal{V}_{inp}|$ bits.
For our running example, Figure~\ref{fig:overview} illustrates the one-hot vector representation of the usage contexts of variables \code{e} and \code{n}.

The one-hot representation of contexts is highly redundant and we next describe how to compress it using a LSTM based sequence auto-encoder.
The auto-encoder can be thought of as two jointly learned functions.
An encoder function $enc : \{0,1\}^{s_c} \rightarrow \R^{E}_{[0,1]}$, which maps the one-hot representation of a context to a real-valued vector of length $E$,
and a decoder function $dec : \R^{E}_{[0,1]} \rightarrow \R^{s_c}_{[0,1]}$ that maps the real-valued vector back to a binary vector of the original length.
The notation $\R^{E}_{[0,1]}$ refers to a vector of length $E$ of real-valued numbers in the range $[0,1]$.
The goal of training the auto-encoder is to minimize the difference between $dec(enc(c)) = \widetilde{c}$ and the original context vector $c$.
The two functions are trained in tandem to minimize this difference.
Once trained, we use the intermediate vector returned by $enc$ for a given context $c$ as the embedding for $c$. 
That is, the auto-encoder compresses the input vector corresponding to a context into an embeddings of length $E$, where $E$ is a hyper-parameter of our approach.

\begin{figure}
    \centering
    \includegraphics[width=\columnwidth]{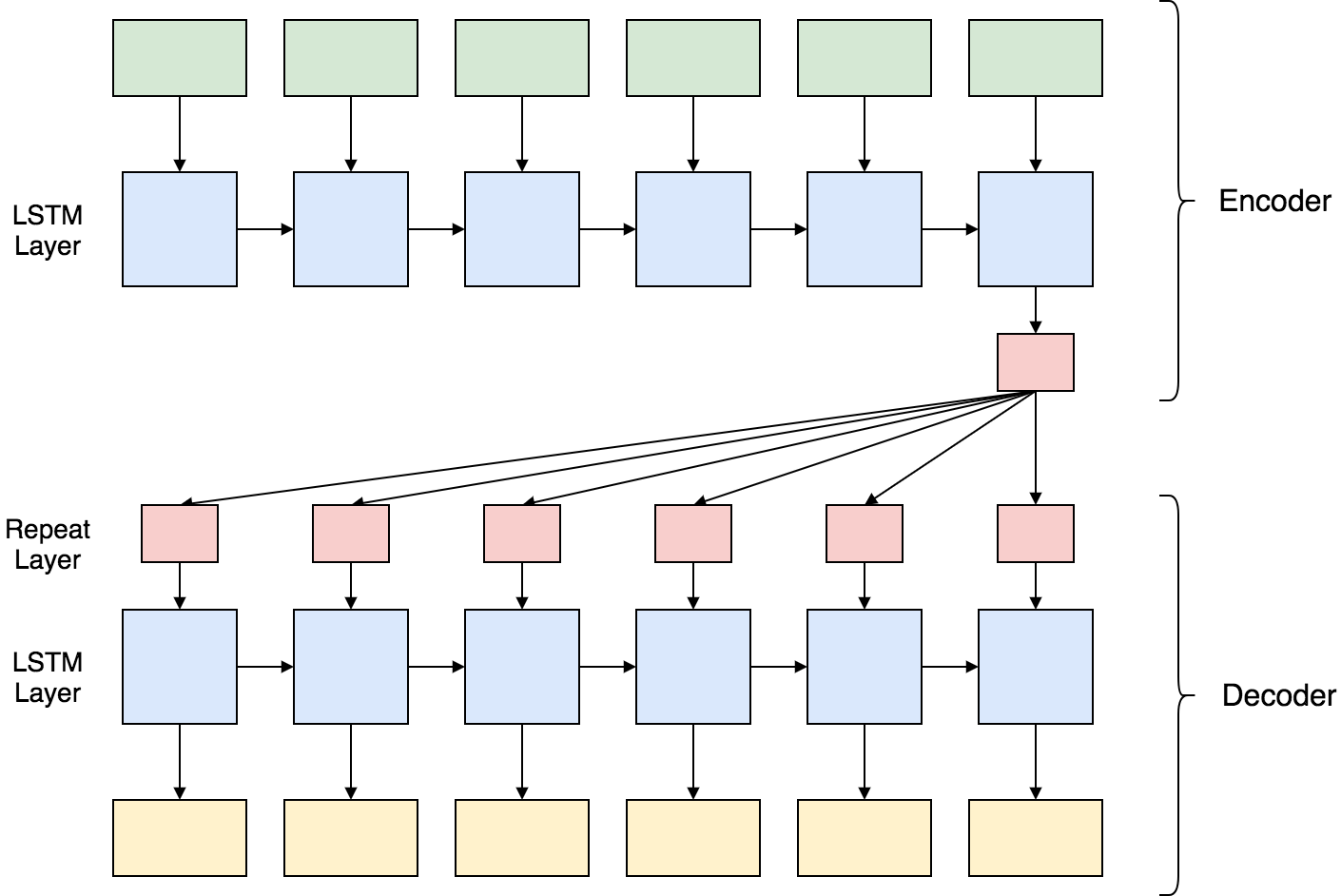}
    \caption{Auto-encoder network that computes an embedding for each context using two RNNs.}
    \label{fig:embedding-network}
\end{figure}

Our implementation of the auto-encoder consist of two jointly trained networks that represent the functions $enc$ and $dec$.
The encoder network consists of a single LSTM layer (a class of recurrent neural network (RNN) models that maintains an internal hidden state and therefore is particularly well-suited to processing sequences of inputs) with an hidden state of size $E$, which also corresponds to the size of the embedding vectors.  The encoder network takes a sequence of $2\cdot q$ one-hot encoded vectors, denoting a context, and produces an embedding vector of length $E$.
The embedding vector is the final hidden state of the LSTM.
We use $E = 80$ as a default value in our implementation.
The decoder network consists of a layer that repeats the input $2 \cdot q$ times (the number of tokens in the context), and a single LSTM layer with a hidden state of size $|\mathcal{V}_{inp}|$.
Figure \ref{fig:embedding-network} provides an illustration of this network.

In summary, we use the encoder component of this network to convert a usage summary for a name $n \in \mathcal{N}$ into a sequence of embeddings:
\begin{align}
\label{eq:embeddings}
\begin{split}
    <enc(c_1), enc(c_2), \cdots , enc(c_l)>\\
         \text{where }\ \mathcal{U}(n) = c_1 \circ c_2 \circ \cdots \circ c_l
\end{split}
\end{align}
For the running example, the ``Dense vector representation'' part of Figure~\ref{fig:overview} shows the real-valued vector that results from concatenating the embeddings of each context in the usage summary of each variable.

\subsection{Predicting Names from Usage Summaries}
\label{sec-predict}

Based on the compactly represented usage summaries, we train a second neural network to predict the name of a variable given its usage summary.
The intuition behind this idea is that the way a variable is used in code implicitly conveys sufficient knowledge about its meaning to predict a suitable name.
We first define an output vocabulary $\mathcal{V}_{out}$ to choose names from.
For our experiments, the vocabulary contains the $60,000$ most frequent names encountered in our code corpus.

We then learn a function $\mathcal{P} : \R^{E \times l}_{[0,1]} \rightarrow \R^{|\mathcal{V}_{out}|}_{[0,1]}$.
Given a sequence of embeddings that represent a usage summary as in Equation~\ref{eq:embeddings}, we first reverse it as suggested in~\cite{Sutskever2014}, so the embeddings constructed out of PAD tokens come first. We then apply the function which yields a probability distribution over the output vocabulary.
This probability distribution can be interpreted as a ranked list of predicted names.

\begin{figure}
    \centering
    \includegraphics[width=\columnwidth]{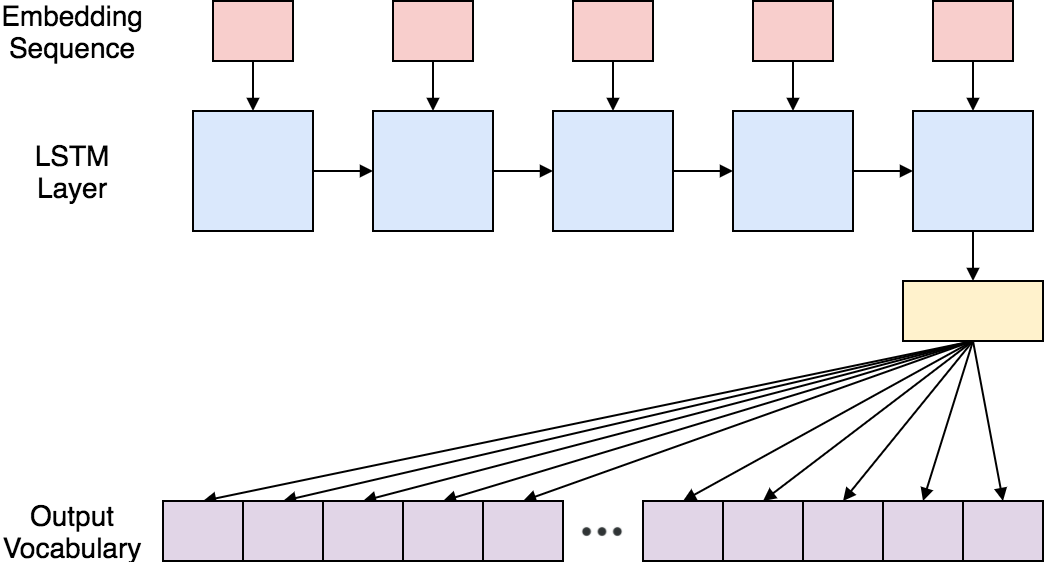}
    \caption{Recurrent neural network to predict likely variable names for a given usage context.}
    \label{fig:prediction-network}
\end{figure}

To learn function $\mathcal{P}$, we use a recurrent neural network (RNN), i.e., a class of neural models that maintains an internal memory and therefore is particularly well-suited to processing sequences of inputs.
The RNN we use consists of a single long short-term memory (LSTM) layer with a hidden state of size $h$ ($h = 3,500$ in our experiments), followed by a softmax layer, which returns a probability distribution. Figure \ref{fig:prediction-network} provides an illustration of this network.

The size of the output vocabulary $\mathcal{V}_{out}$ directly corresponds to the range of names our network can output as a prediction.
That is, the larger the vocabulary, the higher is the accuracy of name recovery.
The trade-off, however, is that the network size increases linearly with $|\mathcal{V}_{out}|$.
The vocabulary size we choose for our experiments ($|\mathcal{V}_{out}| = 60,000$) strikes a balance between performance and precision.

In Figure~\ref{fig:overview}, the lower right part shows the names predicted by \name{} for variables \code{e} and \code{n}.
As in this example, several of the top-ranked names may convey the semantics of the minified variable.

\subsection{Semantics-Preserving Recovery of Names}
\label{sec:recovery}

Our RNN-based predictor outputs a ranked list of possible names for each minified name.
The final step is to map each minified name to a single predicted name.
This mapping must preserve the semantics of the minified program.
Specifically, the same name cannot be mapped as predictions to two different minified variables in the same scope, the predicted name cannot be a keyword, and the predicted name cannot overshadow a name from its parent scope if the name is used in one of its child scopes. 

\begin{algorithm}[t]
\caption{Semantics-Preserving Name Recovery}
\label{alg:nameconflicts}
\begin{algorithmic}[1]
\Procedure{PredictNames}{minifiedCode}
\State recoveredCode $\leftarrow$ copy(minifiedCode)
\State minNames $\leftarrow$ \textsc{GetMinifiedNames}(minifiedCode)
\State pQueue $\leftarrow$ new PriorityQueue()
\\
\ForAll{minName $\in$ minNames}
    \State (pred, prob) $\leftarrow$ \textsc{NextPrediction}(minName)
    \State pQueue.push((prob, minName, pred))
\EndFor
\\
\While{pQueue $\neq$ $\varnothing$}
    \State elem $\leftarrow$ pQueue.pop()
    \State minName $\leftarrow$ elem.minName
    \State pred $\leftarrow$ elem.pred
    \If{\textsc{NoConflicts}(minName, pred)}
        \State Replace minName with pred in recoveredCode
    \Else
        \State (pred, prob) $\leftarrow$ \textsc{NextPrediction}(minName)
        \State pQueue.push((prob, minName, pred))
    \EndIf
\EndWhile
\\
\State \Return recoveredCode
\EndProcedure
\end{algorithmic}
\end{algorithm}

To recover names while respecting these constraints, we compose the ranked predictions for different variables into a single list and then use this list to greedily assign predictions.
Algorithm \ref{alg:nameconflicts} summarizes our approach for recovering names in a consistent and semantics-preserving manner.
The procedure \textsc{PredictNames} takes the minified code as input, makes a copy (line 2), and extracts all the minified names using the \textsc{GetMinifiedNames} procedure (line 3).
The algorithm then initializes a priority queue, which will use the probabilities of name predictions, as returned by the RNN, as the key for sorting in descending order.
The priority queue essentially tracks the minified names yet to be recovered.
Lines 6-9 initialize the priority queue with the top predictions for every name along with their corresponding probabilities.
To this end, we use a procedure \textsc{NextPrediction}, which for a given minified name, returns a pair where the first element is the next best prediction after the previous invocation for the same name, and the second element is its corresponding probability.

Lines 11-21 greedily replaces all minified names with predicted names, as provided by the priority queue, until all names have been replaced, i.e., until the queue becomes empty.
At each iteration of the \texttt{while} loop on line 11, the algorithm pops the element with the highest probability from the priority queue (lines 12-14).
Then, at line 15, a procedure \textsc{NoConflicts} checks whether the algorithm can replace the minified name with the predicted name without creating conflicts.
Specifically, we check whether the name predicted has not already been assigned to a different variable in the same scope, is not a keyword, and does not overshadow a replaced name or global name of the parent scope that is referenced in a child scope.
If the check passes, the algorithm replaces the minified name with the prediction in \texttt{recoveredCode}.
Otherwise, we take the next prediction, and add it to the priority queue.
After the loop ends, the algorithm finally returns the recovered code file.

For our running example, the upper right part of Figure~\ref{fig:overview} shows the code with the names inferred by \name{}. Even though only five out of the local six names are predicted exactly as in the original code, the code is much more readable than its minified version.


\section{Evaluation}
\label{sec:eval}

We have implemented \name{} in Python using Keras\footnote{\url{https://keras.io/}} as the deep learning framework.  The implementation has $397$ lines of Python code.
We now present an experimental evaluation of our approach to demonstrate its effectiveness and applicability. Specifically, we attempt to answer the following research questions:
\begin{itemize}
    \item[\textbf{RQ1:}] How effective is our approach at predicting natural names for minified variables and function names in real-world JavaScript code?
    \item[\textbf{RQ2:}] How does our approach compare to the current state-of-the-art, specifically \jsnice{}~\cite{Raychev2015} and \jsnaughty{}~\cite{Vasilescu2017}?
    \item[\textbf{RQ3:}] Is the approach efficient enough to be practical, and does it scale well to large programs?
\end{itemize}

To answer these questions, we evaluate \name{} with a large corpus of real-world JavaScript files.
The corpus consist of disjoint sets of training files and validation files.
For training, we minify the training files with the popular \uglifyjs{}\footnote{Version 3.1.9, run with the -m (mangler) parameter} tool and then train the approach to recover the original names.
For validation, we give minified versions of the validation files to \name{} and then measure the accuracy of recovering the original, unminified names. Although we use \uglifyjs{} in our experiments, our technique is agnostic of the minifier being used, as long as it does not restructure or remove code. Specifically, our techniques works for all minifiers that only modify variable names, and that remove white-space and other terminals, such as semi-colons and commas.
It is important to note that this accuracy metric gives a \emph{lower-bound} of the effectiveness of any technique, as it is possible that the predicted names may be similar, or even better than the original names.
A qualitative evaluation involving the manual inspection of suggested names is beyond the scope of this work.

\subsection{JavaScript Corpus}
We constructed our training and testing corpus using a publicly available data set\footnote{\url{http://www.srl.inf.ethz.ch/js150.php}} of JavaScript programs~\cite{Raychev2016}. The data set contains 150,000 non-minified JavaScript files: 100,000 files marked for training and 50,000 files marked for testing.
Before our experiments, we cleaned the corpus by the following procedure: We removed 3,150 files that are common between the sets of training and testing files from the training files to ensure that the training and testing data sets are disjoint.
We then removed duplicates from these sets, followed by the removal of files that cannot be processed by \uglifyjs{}. Following the setup of the \jsnice{} artifact, we also remove files that are very large (more than 131,072 characters) or that are already (mostly) minified. This is done to ensure that large or minified files do not skew the results. For fairness, we also remove files that cannot be processed by \jsnice{} or \jsnaughty{} due to implementation-level bugs.
This entire process reduces the number of training and testing files to 64,750 and 33,229, respectively.
A possible alternative to using fixed training and testing sets is k-fold cross validation.
We follow the experimental setup of \jsnice{}~\cite{Raychev2015} and \jsnaughty{}~\cite{Vasilescu2017}, which uses fixed training and testing sets.

We present some statistics about the testing corpus. The number of lines of code in the original source files (excluding comment and blank lines) in the validation corpus ranges from 1 to 7,239 (mean 169.2, median 61). The files contain between 0 to 1,388 unique (per-file, per-scope) local variable names (mean 38.4, median 12), between 1 to 1,401 unique variables (local + global) (mean 45.2, median 18), between 0 to 6,139 (mean 160.1, median 40) usages of local variable names, and between 1 and 6,208 (mean 191.5, median 61) usages of all variable names, i.e., both local and global names.
Across all files, the number of unique local variable names is 1,277,558.
The total number of usages of local variable names is 5,321,106, and the total number of usages of all variable names (global + local) is 6,364,368.

\subsection{Training}
For training \name{}, we build usage summaries for all minified names in the minified files in the training set, and then train our model by using the original names provided by the source maps as the ground-truth.
Across all training files, the total number of unique local variable names is 2,551,118 (i.e., the size of $\mathcal{N}$), which also corresponds to the number of usage summaries used for training.
When collecting unique local variable names, we exclude any names not minified by \uglifyjs{} to preserve the semantics of the code, such as references to built-in functions and global variables.

\subsection{Parameter Selection}

\begin{table}
    \centering
    \setlength{\tabcolsep}{2pt}
    \begin{tabular}{lr}
     \toprule
     Parameter          &  Value \\
     \midrule
     No.\ $q$ of neighbours used on either side in contexts  &  5\\
     No.\ $l$ of contexts used in usage summaries          &  5\\
 Input vocabulary size $|\mathcal{V}_{inp}|$     &  4,096\\
     Output vocabulary size $|\mathcal{V}_{out}|$     &  60,000\\
     Embedding size $E$ &  80\\
     Hidden layer size $h$   &  3,500\\
     \bottomrule
    \end{tabular}
    \caption{Hyperparameters and values for the evaluation.}
    \label{tab:network-params}
\end{table}

The effectiveness and efficiency of \name{} depends on several hyperparameters.
The values of these hyperparameters during the evaluation is provided in Table~\ref{tab:network-params}. 
We found these values to strike a balance between the size of the network and the amount of information used to predict names.
In particular, the vocabulary sizes $|\mathcal{V}_{inp}|$ and $|\mathcal{V}_{out}|$ need to be set carefully. Large vocabularies may capture a lot of information and allow many more names to be predicted, but may severely affect performance.
We select sizes that cover the majority of the tokens in the training set.

We construct both vocabularies using the training data set only, by picking the $|\mathcal{V}_{inp}|$ most frequent tokens across all usage summaries for the input vocabulary, and the $|\mathcal{V}_{out}|$ most frequent names for the output vocabulary.
A key finding is that a relatively small number of tokens in the input vocabulary accounts for a significant fraction of all tokens across all usage summaries.
The reason is that the frequencies of tokens follow a long-tail distribution: Some tokens, such as \texttt{ID} and semi-colons are extremely frequent, whereas many other tokens, such as application-specific literal values, occur only rarely. 
The situation is similar for the output vocabulary, where a relatively small set of popular identifier names cover the large majority of all occurrences of identifiers.

\begin{table}[]
    \centering
    \begin{tabular}{rp{9em}p{7.5em}}
    \toprule
        $|\mathcal{V}_{inp}|$ & Percentage of unique tokens covered & Percentage of tokens covered\\
        \midrule
        500    & 0.09    & 93.39\\
        1,000  & 0.19   & 94.53\\
        2,000  & 0.38   & 95.46 \\
        3,000  & 0.56   & 95.92 \\
        4,000  & 0.75   & 96.23 \\
        \textbf{4,096}  & \textbf{0.77}   & \textbf{96.26} \\
        5,000  & 0.94   & 96.46 \\
        6,000  & 1.13   & 96.64 \\
        7,000  & 1.32   & 96.79 \\
        8,000  & 1.50   & 96.91 \\
        9,000  & 1.69   & 97.02 \\
        10,000 & 1.88   & 97.11 \\
    \bottomrule
    \end{tabular}
    \caption{Impact of input vocabulary size $|\mathcal{V}_{inp}|$. The bold line is our default setting.}
    \label{tab:inp-vocab-discussion}
\end{table}

\begin{table}[]
    \centering
    \begin{tabular}{rp{9em}p{7.5em}}
    \toprule
        $|\mathcal{V}_{out}|$ & Percentage of unique names covered & Percentage of names covered\\
        \midrule
        1,000   & 0.40    & 63.19\\
        5,000   & 1.99    & 75.07\\
        10,000  & 3.97    & 79.48\\
        20,000  & 7.95    & 83.82\\
        30,000  & 11.92   & 86.38\\
        40,000  & 15.89   & 88.16\\
        50,000  & 19.87   & 89.56\\
        \textbf{60,000}   & \textbf{23.84}   & \textbf{90.74}\\
        70,000  & 27.81   & 91.62\\
        80,000  & 31.79   & 92.41\\
        90,000  & 35.76   & 93.19\\
        100,000 & 39.74   & 93.98\\
    \bottomrule
    \end{tabular}
    \caption{Impact of output vocabulary size $|\mathcal{V}_{out}|$.  The bold line is our default setting.}
    \label{tab:out-vocab-discussion}
\end{table}

Tables~\ref{tab:inp-vocab-discussion} and~\ref{tab:out-vocab-discussion} show the impact of different vocabulary sizes. The middle column in Table~\ref{tab:inp-vocab-discussion} compares the relative size of the input vocabulary to the number of unique tokens (531,943) seen across all usage summaries in the training set.
Note that tokens corresponding to minified names are replaced by the special \texttt{ID} token in usage summaries.
The right column in Table~\ref{tab:inp-vocab-discussion} shows the percentage of all tokens, seen across all usage summaries, that are present in the input vocabulary. The entries in Table~\ref{tab:inp-vocab-discussion} suggest that the input vocabulary, albeit very small in comparison to the set of all \emph{unique} tokens, covers a very high fraction of \emph{all} tokens (96.26\% of the tokens from a total of 127,555,900 tokens for $|\mathcal{V}_{inp}| = 4{,}096$).

Table~\ref{tab:out-vocab-discussion} paints a similar picture for the output vocabulary. The middle column compares the relative size of the output vocabulary to the number of unique non-minified names (251,663) seen in our training set, and the right column shows the percentage of all names covered by the output vocabulary. The entries in bold correspond to the size we chose ($|\mathcal{V}_{out}| = 60{,}000$). The conclusion is the same as for the input vocabulary: The output vocabulary is small but sufficiently rich (recognizing 90.74\% of the names from a total of 2,551,118 local variables) to perform naming tasks satisfactorily.

\begin{figure*}
    \centering
    \begin{minipage}[t]{0.4\textwidth}
        \centering
        \begin{tabular}{p{7.5em}|p{5em}p{7em}}
            \toprule
            & Predict local names only & Predict all (local + global) names \\
            \midrule
            Count each variable once & Local-Once & All-Once \\[1em]
            
            Count each occurrence of a variable & Local-Repeat & All-Repeat \\
            \bottomrule
        \end{tabular}
        \captionof{figure}{Accuracy metrics used for the evaluation.}
        \label{tab:metrics}
    \end{minipage}
    \hspace{2em}
    \begin{minipage}[t]{0.55\textwidth}
        \centering
        \vspace{-4.5em}        
        \setlength{\tabcolsep}{2pt}
        \begin{tabular}{lrrrrr}
        \toprule
        Metric         & \name{}    & \jsnice{} & \jsnaughty{} & \jsnaughty{}$^\infty$ & Baseline\\
        \midrule
        Local-Once     & 47.5\%     & 48.3\%    & 39.4\%    & 55.3\%       & 0.0\%\\
        Local-Repeat   & 49.8\%     & 55.3\%    & 41.3\%    & 59.2\%       & 0.0\% \\
        All-Once       & 55.4\%     & 56.0\%    & 47.7\%    & 61.9\%       & 15.0\%  \\
        All-Repeat     & 58.1\%     & 62.6\%    & 49.3\%    & 65.8\%       & 16.4\% \\
        \bottomrule
        \end{tabular}
    \captionof{figure}{Comparison of accuracy (metrics described in Section \ref{sec-accuracy-metrics}) for \name{}, \jsnice{}, and \jsnaughty{}. Baseline represents a tool which does not predict anything. \jsnaughty{}$^\infty$ refers to the \jsnaughty{} without imposing any time limit.}
        \label{tab:accuracy-metrics-1-2-no-timeout}
    \end{minipage}
\end{figure*}

\subsection{Setup for \jsnice{} and \jsnaughty{}}

To compare \name{} with the state-of-the-art, we train \jsnice{} and \jsnaughty{} on the same training set as ours, enabling an apples-to-apples comparison. We use the \jsnice{} artifact\footnote{\url{http://files.srl.inf.ethz.ch/jsniceartifact/index.html}}, and use the same parameters and arguments as suggested in the accompanying README file.
For prediction tasks, we again use the same command line arguments as provided in the README, with an additional parameter to produce a source map from the minified file to the file recovered by \jsnice{}.
We then use this map to compute the accuracy.

For \jsnaughty{}, we trained both their translation and language models, as well as the Nice2Predict framework\footnote{\url{http://www.nice2predict.org/}}~\cite{Raychev2015} on our training set by following instructions provided by the authors. It is important to train a Nice2Predict instance because \jsnaughty{} combines results from both their models and Nice2Predict to make the final prediction.

By default, we run each of three tools with a time limit of 10 minutes for processing a file.
If a timeout occurs, we assume that the respective tool failed to recover any names in this file.
Imposing a time limit is reasonable for de-minification tools for two reasons.
First, developers use such tools to save time while inspecting code, and waiting for a tool to finish defeats this purpose.
Second, such tools usually have a web-interface (both \jsnice{} and \jsnaughty{} have a web interface) and response times are critical to satisfactorily serve a large number of users.
We have not yet released a web interface for \name{} to maintain anonymity, but will do so once this work is accepted.
In practice, the 10-minute time limit affects only \jsnaughty{}.
For a full comparison, we also run \jsnaughty{} without any time limit, where it takes over 40 minutes for some files.

\subsection{Evaluation Criteria}\label{sec-accuracy-metrics}

The evaluation criterion for assessing the effectiveness of the approaches is accuracy, i.e, the ability to recover the original names from minified files.
To measure accuracy, we run \name{}, \jsnice{}, and \jsnaughty{} on each minified file in the validation corpus, and extract a mapping between the minified names and the corresponding predictions made by the tools. We then combine this mapping with the source maps produced by \uglifyjs{} to define a mapping between the original names and the predicted names.
Finally, we measure accuracy by computing the percentage of original names that a tool recovers correctly.

There are four variants of the ``accuracy'' metric, which differ in subtle ways.
The metrics differ in two dimensions, as illustrated in Table~\ref{tab:metrics}.
On the one hand, we can measure accuracy either for all variables and functions in the code, which includes global names, or only consider local names.
Since minifying global names in a file may break the semantics of the code, minifiers, such as \uglifyjs{}, do not modify these names.
Hence, the task of recovering global names is trivial, as they are not minified at all.
Arguably, both definitions of accuracy make sense, and therefore we consider both of them.

On the other hand, we can measure accuracy either per variable or per usage of a variable.
For example, if a local variable \code{foo} is referenced three times in the same scope, then the per variable metric counts the prediction for \code{foo} once, whereas the per usage metric counts the prediction three times.
If the variable \code{foo} appears twice in different scopes in the same file, then the per variable metric counts the prediction for \code{foo} twice.
Again, both definitions make some sense, so we here consider both.

Together, these two dimensions yield four accuracy metrics, Local-Once, All-Once, Local-Repeat, All-Repeat, as shown in Table~\ref{tab:metrics}.
Metric Local-Repeat can be seen as a weighted computation of metric Local-Once, in that correctly predicting a name for a more frequently used variable would give a higher score. Metric All-Repeat (used by \jsnice{}~\cite{Raychev2015}) allows us to directly gauge the similarity between the original, non-minified file and the recovered file.
As a trivial baseline, we also compute a \emph{baseline accuracy}, which gives the percentage of global names only. That is, the baseline accuracy effectively represents the accuracy of a tool that does not predict any names.

\subsection{Results}

\subsubsection{Accuracy}\label{sec-accuracy-analysis}

Addressing RQ1 and RQ2, Table~\ref{tab:accuracy-metrics-1-2-no-timeout} provides the four accuracy metrics for all the three techniques.
The Local-Once accuracy of \name{}, i.e., when only unique occurrences of local variable names are considered, is 47.5\%, only 0.8\% lower than \jsnice{}'s accuracy of 48.3\%, and 8.1\% higher than \jsnaughty{}'s accuracy of 39.4\%.
When not imposing any time limit, \jsnaughty{} performs better and reaches an accuracy of 55.3\%.
However, this increase comes at a high cost on efficiency (Section~\ref{sec-timing-analysis}), which significantly reduces the practicability of the tool.
Setting a stricter time limit of only one minute, the Local-Once accuracy of \jsnaughty{} drops even further to only 8.9\%.
The main reason for \jsnaughty{}'s sharp drop in accuracy when imposing a time limit is that it suffers from an inherent scalability problem, which the authors, on their GitHub page, attribute to the size of phrase-tables used in predicting names.
For the experiments reported in the \jsnaughty{} paper~\cite{Vasilescu2017}, the tool was run only on files with 100 lines or less.

\subsubsection{Detailed Comparison}

\begin{figure}
    \begin{subfigure}{\linewidth}
        \centering
        \includegraphics[width=.8\columnwidth]{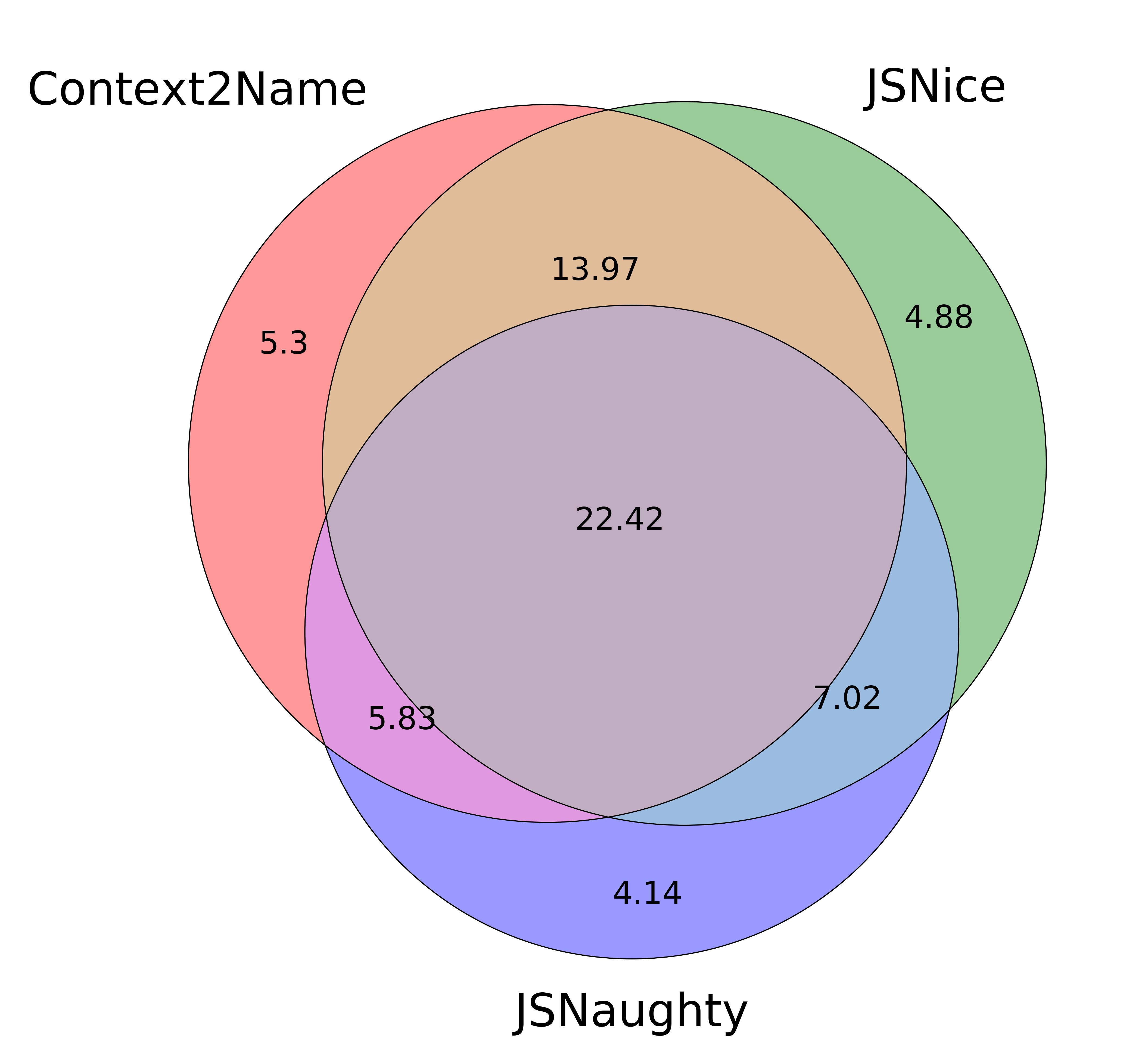}
        \caption{Comparison of predictions by \name{}, \jsnice{} and \jsnaughty{} when a time limit of 10 minutes is imposed.}
        \label{fig:venn-full}
    \end{subfigure}
    
    \begin{subfigure}{\linewidth}
        \centering
        \includegraphics[width=.8\columnwidth]{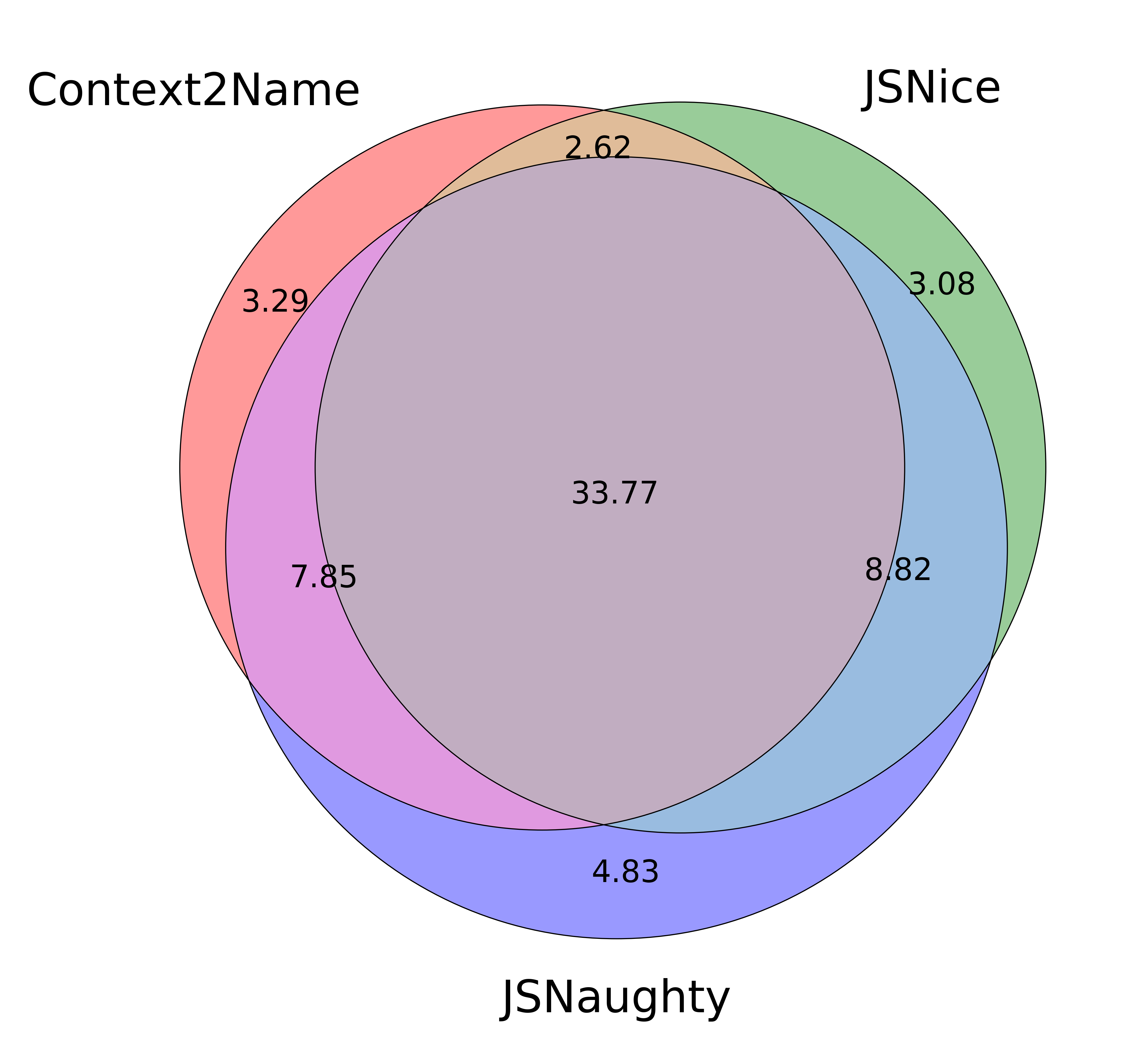}
        \caption{Comparison of predictions by \name{}, \jsnice{} and \jsnaughty{} when no time limit is imposed.}
        \label{fig:venn-partial}
    \end{subfigure}
\end{figure}

We also analyze the relationships between the sets of names recovered by the three approaches.
In the Venn diagram shown in Figure~\ref{fig:venn-full}, each approach is represented by a colored circle.
The percentages inside the colored regions represent the percentage of local minified names for which the original names are recovered correctly (i.e., based on the Local-Once metric). The sizes of the regions also reflect this percentage. It is easy to verify that the sum of percentages inside every circle is equal to the Local-Once accuracies presented in Table~\ref{tab:accuracy-metrics-1-2-no-timeout}.
Across all files, 5.3\% of the total number of unique local variable names are recovered correctly only by \name{}, 4.88\% only by \jsnice{} and 4.14\% only by \jsnaughty{}. 13.97\% of the names are recovered only by \name{} and \jsnice{} and 22.42\% of the names were recovered by all three tools.
Figure~\ref{fig:venn-partial} shows the results for running \jsnaughty{} without any time limit.
Again, each tool recovers some names that the other tools miss.
Overall, the two diagrams suggest that all tools are complementary to each other to some extent.
Thus, it may be possible to build a tool that combines all the three approaches to yield accuracies upwards of 60\%.

\subsubsection{Efficiency}\label{sec-timing-analysis}

To address RQ3 about the efficiency and scalability of \name{}, we measure the time needed for predicting names.
We performed our experiments on a 32-core machine with four 2.40 GhZ Intel Xeon processors running Ubuntu 16.04.1 64-bit, with 256GB RAM.
We trained and hosted our model on a separate machine with one 4.20 GhZ Intel i7 processor, with one Nvidia 1080Ti GPU, running Ubuntu 16.04.1 64-bit with 48 GB RAM.
The embedding and prediction networks are trained separately, both for 5 epochs, and the time to train them was 1.5 days and 3 days, respectively, which is a one-time effort.
For prediction tasks, we developed a client running on the first machine that queries the model hosted on the second. Our timing therefore is the sum of processing time on the first machine, and the time taken to query the model loaded on the GPU. 

\begin{table}
    \centering
    \setlength{\tabcolsep}{2pt}
    \begin{tabular}{lrrrr}
    \toprule
    Name          & Min. (ms)  & Max. (ms)   & Mean (ms)    & Median (ms) \\
    \midrule
    \name{}       & 0.3       & 2,557.2      & 110.7        & 52.0 \\
    \jsnice{}     & 7.0       & 13,151.0     & 270.3        & 73.0 \\
    \jsnaughty{}  & 1.5       & 2,489,076.2  & 64,962.2     & 20,043.0 \\
    \bottomrule
    \end{tabular}
    \caption{Timing statistics for \name{}, \jsnice{} and \jsnaughty{}, computed per file in the testing set. The columns shows the minimum, maximum, mean, and median running time in milliseconds. No time-out is imposed on any of the tools.}
    \label{tab:timing}
\end{table}

Table~\ref{tab:timing} shows the per-file timing statistics. Both \name{} and \jsnice{} perform comparably and are able to process most files in under half a second, making the techniques good candidates for online interactive tools. \jsnaughty{} takes an average of 65 seconds.
Thus, the improvement in accuracy of \jsnaughty{} comes at a very significant penalty in efficiency, making it less suitable to be deployed in an interactive setting.

\medskip
\noindent
Overall, the results show that \name{} performs comparably to the existing techniques with respect to accuracy, and that it outperforms the existing techniques, in particular \jsnaughty{}, with respect to running time.
Moreover, \name{} complements existing techniques by predicting 5.3\% of all names that are missed by both existing tools.
These results are particularly remarkable given that \name{} does not rely on any kind of manual and language-specific feature engineering, making it straightforward to apply the approach to another programming language.


\section{Related Work}
\subsection{Obfuscation}

Program obfuscation has applications in protection of intellectual property~\cite{Drape2009}, resistance against reverse-engineering and software tampering~\cite{Wang2000, Balachandran2013, Junod2015} as well as watermarking~\cite{Collberg2002}. An obfuscator is basically a compiler, that takes an input program, and outputs a semantically equivalent program that is mostly unintelligible to a third party without access to the source. Barak et al.~\cite{Barak2012} showed that obfuscation in general is not realizable. Nonetheless a large body of work has been published on new techniques for obfuscation~\cite{collberg1997taxonomy,DBLP:conf/icse/LiuSSJGS17}, as it has proved to be practically useful.

Obfuscation is an attractive option in the domain of JavaScript programs as the code is shipped as source, allowing anyone to view and download the original code of the author. But excessive obfuscation can be detrimental to performance w.r.t. bandwidth usage and execution time. Obfuscators that increase the size of the program, or make it significantly slower are not particularly useful. Compatibility is also an issue, as many JavaScript programs rely on external APIs whose usage has to appear in the clear. Minifiers, such as \uglifyjs{}, are an excellent compromise, as the resultant programs are much smaller, and variable renaming is a sufficient deterrence for most adversaries. 

\subsection{Deobfuscation}
Deobfuscation techniques attempt to uncover various aspects of the semantics of the program, which has applications in reverse engineering and malware analysis. Most of the proposed techniques rely on static and dynamic analyses~\cite{Moser2007, Udupa2005, Christodorescu2003}, which are more adept at uncovering semantic information about the obfuscated program. Instead, our work focuses on producing friendlier summarizations in the form of variable names, which is useful in the domain of JavaScript programs. Statistical techniques that relate the semantics to the names used in the program are expected to perform better at this task.

\subsection{Probabilistic Models for Code}
Machine learning has been used in various domains such as natural language, computer vision with tremendous success. These successes have created a lot of interest in applying machine learning to programs, targeting a variety of development tasks such as code completion~\cite{Raychev2014,Raychev2016a, Bielik2016}, fixing syntactic errors~\cite{Gupta2017, Bhatia2016}, generating inputs for fuzz testing of programs~\cite{Godefroid2017, Patra2016, Liu2017}, and generating programs or short code snippets to assist developers or test compilers~\cite{Amodio2017, Shu2017}. Deep learning techniques in particular have been used for code synthesis~\cite{Balog2016}, detecting clones of code fragments~\cite{White2016}, and malware detection and signature generation~\cite{David2015}. The core idea behind all these techniques is to exploit the abundance of available source code by mining useful patterns and correlating them to the value of the desired property. In our case, we try to correlate a name to a context comprised of its surrounding tokens, which may include identifiers that have not been minified (global names).

The two closest existing techniques are \jsnice{}~\cite{Raychev2015} and \jsnaughty{}~\cite{Vasilescu2017}, and we show in Section~\ref{sec:eval} that our approach outperforms both.
Conceptually, our work differs from \jsnice{} by not relying on any program analysis to extract features that may be relevant for predicting variable names, but to instead let neural networks decide which parts of the syntactic context matter.
A practical benefit of this design decision is that \name{} is less language-dependent and can be adapted to another language easily.
Compared to \jsnaughty{}, an important difference concerns the efficiency and scalability to larger files.
As discussed in Section~\ref{sec-accuracy-analysis}, \jsnaughty{} takes very long for files with only a few hundreds of lines code, which is why the evaluation of \jsnaughty{} in~\cite{Vasilescu2017} only considers files with up to 100 lines.

Statistical techniques which also use identifiers have been used to tackle a number of problems. \jsnice{}~\cite{Raychev2015}, in addition to predicting names, also predicts types for variables in JavaScript. We believe our approach of using sequences of neighbouring lexical tokens can be extended to type prediction as well. Another approach~\cite{Allamanis2016} tries to predict concise yet descriptive names that summarize code snippets, which can then effectively serve as method names. \textsc{Naturalize}~\cite{Allamanis2014} is a framework that enforces a consistent naming style by providing better names. We expect our number of incorrect predictions to go down after using \textsc{Naturalize} on our testing corpus, as the number of ``surprising'' and inconsistent names would decrease.

Sequences of lexical tokens are also used by the \textsc{SmartPaste} framework~\cite{Allamanis2017} which is designed to assist a developer in extending an existing code base with snippets, by attempting to align the variables in the added snippet with the rest of the code. \textsc{SmartPaste} uses the lexical tokens around a variables in the snippet to compute the probability of it being replaced by some variable in the rest of the code. In addition to using lexically preceding tokens, they also use data flow relations to compute the relevant tokens surrounding the use of a variable. We expect to benefit from using data flow analysis as well.

Embeddings are popular in machine learning-based natural language processing, where they abstract words into vectors~\cite{Mikolov2013a}.
Beyond pure natural language texts, embeddings for code elements have been proposed, e.g., for API elements~\cite{Nguyen2017} and for terms that appear both in code and in natural language documentation~\cite{Ye2016}.
Our embeddings differ from these approaches by serving a different purpose, namely to compress usage contexts while preserving their important features, instead of reasoning about individual program elements.

\section{Conclusion}

This paper addresses the problem of recovering meaningful identifier names in code that has been obfuscated by replacing all local names with short, arbitrary, and meaningless names.
We present a deep learning-based approach that exploits and learns from the large amount of available code.
The key idea is to predict from the syntactic usage context of a variable what meaning the variable has and to then assign a suitable name.
To this end, we combine a lightweight static analysis, an auto-encoder neural network, and a recurrent neural network into a fully automated tool.
An evaluation on a set of 150,000 JavaScript files shows that \name{} successfully predicts 47.5\% of all
minified identifiers while taking only 2.9 milliseconds on average to
predict a name. A comparison with the state-of-the-art tools JSNice
and JSNaughty shows that our approach performs comparably in
terms of accuracy while improving in terms of efficiency. In fact \name{} is complementary to the aforementioned tools in that it predicts 5.3\% additional identifiers missed by them. 
In conclusion, our work contributes a practical tool to help developers understand minified code and shows the power of deep learning to reason about code.
Because the approach makes few assumptions about the analyzed programming language and uses a very lightweight, token-based program analysis, we believe it will be easily applicable to other languages.

\begin{acks}
This work was supported in part by NSF grants CCF-1409872 and CCF-1423645, by a gift from Fujitsu Laboratories of America, Inc., by the German Federal Ministry of Education and Research and by the Hessian Ministry of Science and the Arts within CRISP, and by the German Research Foundation
within the Emmy Noether project ConcSys.
\end{acks}

\bibliography{references}



\end{document}